\documentclass[pra,twocolumn,superscriptaddress]{revtex4-1}
\usepackage{amsmath,amsfonts,url}
\usepackage{graphicx}

\usepackage{theorem}

\theorembodyfont{\normalfont}
\newtheorem{theorem}{Theorem}
\newtheorem{definition}{Definition}
\newtheorem{corollary}{Corollary}
\newtheorem{conjecture}{Conjecture}
\newtheorem{lemma}{Lemma}
\newtheorem{fact}{Fact}
\newcommand{\bit}{\{0,1\}}
\newcommand{\bits}[1]{\{0,1\}^{#1}}

\begin{document}
\begin{flushright}
YITP-19-02
\end{flushright}

\title{Fine-grained quantum computational supremacy}
\author{Tomoyuki Morimae}
\email{tomoyuki.morimae@yukawa.kyoto-u.ac.jp}
\affiliation{Yukawa Institute for Theoretical Physics,
Kyoto University, Kitashirakawa Oiwakecho, Sakyoku, Kyoto
606-8502, Japan}
\affiliation{JST, PRESTO, 4-1-8 Honcho, Kawaguchi, Saitama,
332-0012, Japan}
\author{Suguru Tamaki}
\email{tamak@sis.u-hyogo.ac.jp}
\affiliation{Graduate School of Informatics, Kyoto University,
Yoshida-Honmachi, Sakyoku, Kyoto 606-8501, Japan}
\affiliation{School of Social Information Science,
University of Hyogo, 8-2-1, Gakuennishi-machi,
Nishi-ku, Kobe, Hyogo 651-2197, Japan}
\date{\today}

\begin{abstract}
Output probability distributions of
several sub-universal quantum computing models 
cannot be classically efficiently sampled
unless some unlikely consequences occur in
classical complexity theory, such as the collapse of the 
polynomial-time hierarchy.
These results, so called quantum supremacy, however,
do not rule out possibilities of 
super-polynomial-time classical simulations.
In this paper,
we study ``fine-grained" version of quantum supremacy
that excludes some exponential-time classical simulations.
First, we focus on two sub-universal models, namely,
the one-clean-qubit model (or the DQC1 model)
and the HC1Q model.
Assuming certain conjectures in fine-grained complexity theory,
we show that 
for any $a>0$
output probability distributions of these models
cannot be classically sampled within a constant multiplicative error
and in $2^{(1-a)N+o(N)}$ time, where $N$ is the number of qubits.  
Next, we consider universal quantum computing.
For example, we consider
quantum computing over Clifford and $T$ gates, 
and show that
under another fine-grained complexity
conjecture, 
output probability distributions
of
Clifford-$T$ quantum computing 
cannot be classically sampled
in $2^{o(t)}$ time 
within a constant multiplicative error, 
where $t$ is the number of $T$ gates.
\end{abstract}

\maketitle
\section{Introduction}
Quantum computing is believed to outperform classical
computing. In fact, quantum advantages  
have been shown in terms of, for example,
communication complexity~\cite{comm1,comm2,comm3,comm4}.
Regarding time complexity,
however,
the ultimate goal, ${\rm BQP}\neq{\rm BPP}$, seems to be extremely hard
to show because of the ${\rm P}\neq{\rm PSPACE}$ barrier~\cite{complexity_zoo}.

Mainly three types of approaches exist
to demonstrate quantum speedups over classical computing.
The first approach is to construct quantum algorithms
that are faster than known best classical algorithms.
For example, quantum computing can do
factoring~\cite{Shor}
and simulations of quantum many-body dynamics~\cite{Iulia}, etc.
faster than known best classical algorithms.
One disadvantage of this approach is, however,
that classical best algorithms could be updated, e.g., Ref.~\cite{Tang}.
The second approach is to study query complexity.
Quantum computing has been shown to require fewer oracle queries than
classical computing
for some problems~\cite{Simon,Grover}.
Query complexity is, however, not necessarily equal to
time complexity.

The third approach, which has been actively studied
recently, is to reduce classical simulatabilities
of quantum computing
(in terms of sampling)
to certain unlikely collapses of conjectures in classical
computational complexity theory,
such as the infiniteness of the polynomial-time hierarchy.
Several sub-universal quantum computing models have been proposed,
such as the depth-four circuits~\cite{TD},
the Boson Sampling model~\cite{BS},
the IQP model~\cite{IQP1,IQP2},
the one-clean qubit model (or the DQC1 model)~\cite{KL,MFF,M,Kobayashi,KobayashiICALP},
the random circuit model~\cite{random,random2,random3},
and the HC1Q model~\cite{HC1Q}.
(Definitions of the IQP model, DQC1 model,
and HC1Q model are given later.)
It has been shown that if output probability distributions of
these models are classically 
sampled in polynomial-time within a multiplicative error,  
then the polynomial-time hierarchy collapses
to the second level.
Here, we define the multiplicative error sampling as follows.

\begin{definition}
We say that the acceptance probability $p_{acc}$ of a quantum computer
is
classically sampled in time $T$ within a multiplicative error $\epsilon$
if there exists a classical probabilistic 
algorithm that accepts with probability $q_{acc}$ in time $T$ 
such that 
\begin{eqnarray}
|p_{acc}-q_{acc}|\le\epsilon p_{acc}.
\end{eqnarray}
\end{definition}
The polynomial-time hierarchy is not believed to collapse
in classical complexity theory, and therefore if we conjecture
the infiniteness of the polynomial-time hierarchy,
the reductions suggest that
quantum computing is faster than classical computing
for the sampling task. This type of quantum advantage is called
quantum supremacy.

This ``traditional" quantum supremacy prohibits classical 
{\it polynomial-time}
sampling, 
but 
any possibility of 
super-polynomial-time 
classical sampling is not excluded.
For example,
a sub-universal quantum computer with $N$ qubits cannot be classically sampled
in $poly(N)$ time, but 
it could be classically sampled in, say, 
$N^{\log^*N}$ time,
where 
$
\log^*N\equiv
1+\log^*(\log N)
$
is the iterated logarithm.
For small $N$ that the current near-term
devices (so called NISQ devices) aim at,
$N^{\log^*N}$-time classical simulation could be tractable.
Hence 
more ``fine-grained" versions of quantum supremacy
that exclude super-polynomial-time classical simulations
are necessary.

To show such fine-grained quantum supremacy, 
``traditional" complexity classes, such as P, NP, and the polynomial-time
hierarchy, do not seem to
be useful because only differences between
polynomial or exponential are considered. In classical computer science,
so called fine-grained complexity theory has emerged recently, where
exact complexities of problems are studied.
Several conjectures, such as SETH, 3SUM, and APSP, are introduced,
and lowerbounds of complexities of many problems have been 
derived~\cite{FGreview}.
(The definition of SETH is given later. We do not use 3SUM and APSP in
this paper. For their definitions, see Ref.~\cite{FGreview}.)
Can we show impossibilities of some super-polynomial-time
classical simulations based on these fine-grained conjectures?

Dalzell et al. showed that
under some fine-grained complexity conjectures,
output probability distributions of
the IQP model, the Boson sampling model, 
and the QAOA model cannot be classically sampled within a constant
multiplicative error in certain exponential time~\cite{Dalzell,DalzellPhD}.
Huang et al. showed impossibilities of
strong simulations (i.e., exact computations of output probability
distributions) in some exponential time
assuming the exponential time hypothesis (ETH)~\cite{Huang,Huang2}.
(The definition of ETH is given later.
Relations between our results and theirs are discussed in detail
in Secs.~\ref{sec:IQP} and~\ref{sec:BS}.)

The purpose of this paper is to study fine-grained quantum supremacy
for other various models including
the DQC1 model, the HC1Q model, and some universal model such as
the Clifford-$T$ model.
We show that based on some fine-grained complexity conjectures,
these models cannot be classically sampled within a constant multiplicative
error in certain exponential time.
This paper is organized as follows. In the next section, 
Sec.~\ref{sec:conjectures}, we provide
conjectures we use.
In Sec.~\ref{sec:DQC1}, we show fine-grained quantum supremacy for
the DQC1 model.
In Sec.~\ref{sec:HC1Q}, we show fine-grained quantum supremacy for
the HC1Q model.
In Sec.~\ref{sec:univ}, we study fine-grained quantum supremacy for
universal models such as the Clifford-$T$ model.
In Sec.~\ref{sec:stabilizer}, we study the stabilizer rank that
is closely related to the fine-grained quantum supremacy of
the Clifford-$T$ model.
Finally, in Sec.~\ref{sec:discussion}, we give some discussions.
All proofs are given in Sec.~\ref{sec:proofs}.

\section{Conjectures}
\label{sec:conjectures}
In this section, we introduce conjectures we use.
(It is summarized in Table~\ref{table}.)
Our conjectures are, in some sense, related to 
the exponential-time hypothesis (ETH), the strong-exponential-time
hypothesis (SETH), and the non-deterministic strong-exponential-time hypothesis
(NSETH).
Here, ETH, SETH, and NSETH are defined as follows.
(Note that the original ETH is about $2^{\Omega(n)}$-time scaling,
but it is equivalent to
Conjecture~\ref{conjecture:ETH}~\cite{IPZ01}.)

\begin{conjecture}[ETH]
\label{conjecture:ETH}
Any (classical) deterministic algorithm
that decides whether $\# f>0$ or $\# f=0$
given
(a description of) a 3-CNF with $m$ clauses,
$f:\{0,1\}^n\to\{0,1\}$,
needs $2^{\Omega(m)}$ time.
Here,
\begin{eqnarray}
\#f\equiv\sum_{x\in\{0,1\}^n}f(x).
\end{eqnarray}
\end{conjecture}

\begin{conjecture}[SETH]
\label{conjecture:SETH}
Let $A$ be any (classical) deterministic $T(n)$-time
algorithm such that the following holds:
given (a description of) a 
CNF, $f:\{0,1\}^n\to\{0,1\}$, with at most $cn$ clauses,
$A$ accepts if $\#f>0$ and rejects if $\#f=0$,
where 
\begin{eqnarray}
\#f\equiv\sum_{x\in\{0,1\}^n}f(x).
\end{eqnarray}
Then, for any constant $a>0$, there exists a constant $c>0$
such that $T(n)>2^{(1-a)n}$ holds for
infinitely many $n$.
\end{conjecture}

\begin{conjecture}[NSETH]
\label{conjecture:NSETH}
Let $A$ be any non-deterministic
$T(n)$-time algorithm such that
the following holds:
given (a description of)
a polynomial-size Boolean circuit 
$f:\{0,1\}^n\to\{0,1\}$, $A$ accepts if
$\#f=0$ and rejects if $\#f>0$, where
\begin{eqnarray}
\#f\equiv\sum_{x\in\{0,1\}^n}f(x).
\end{eqnarray}
Then, for any constant $a>0$,
$T(n)>2^{(1-a)n}$
holds for infinitely many $n$.
\end{conjecture}

Note that usually $k$-CNFs, not polynomial-size Boolean circuits, are
considered in NSETH. Here, we consider the more general one.

Now we introduce our conjectures.
First let us consider the following three conjectures.
\begin{conjecture}
\label{conjecture:general}
Let $A$ be any non-deterministic $T(n)$-time algorithm such that
the following holds: given (a description of) a polynomial-size
Boolean circuit $f:\{0,1\}^n\to\{0,1\}$,
$A$ accepts if $gap(f)\neq0$ and rejects if $gap(f)=0$, where
\begin{eqnarray}
gap(f)\equiv \sum_{x\in\{0,1\}^n}(-1)^{f(x)}.
\end{eqnarray}
Then, for any constant $a>0$, $T(n)> 2^{(1-a)n}$ holds for infinitely
many $n$. 
\end{conjecture}
\begin{conjecture}
\label{conjecture:NC}
It is the same as Conjecture~\ref{conjecture:general}
except that
the
Boolean circuit $f$ is of logarithmic depth.
\end{conjecture}
\begin{conjecture}
\label{conjecture:reversible}
Let $A$ be any non-deterministic $T(n)$-time algorithm such that
the following holds: given (a description of) a polynomial-size
classical reversible circuit $C:\{0,1\}^{n+\xi}\to\{0,1\}^{n+\xi}$ 
with $\xi\in o(n)$
that consists
of only NOT and TOFFOLI,
$A$ accepts if $gap(C)\neq0$ and rejects if $gap(C)=0$, where
\begin{eqnarray}
gap(C)\equiv \sum_{x\in\{0,1\}^n}(-1)^{C_{n+\xi}(x0^\xi)},
\end{eqnarray}
$x0^\xi$ is the bit string composed of $x$ and $\xi$ 0s,
and $C_{n+\xi}(x0^\xi)\in\{0,1\}$ is the last bit of 
$C(x0^\xi)\in\{0,1\}^{n+\xi}$. 
Then, for any constant $a>0$, $T(n)> 2^{(1-a)n}$ holds for infinitely
many $n$. 
\end{conjecture}

One difference between these three conjectures and SETH is that
instead of deterministic algorithms, non-deterministic
algorithms are considered. The other difference is that our conjectures
consider gap functions, while SETH considers SAT problems.
These three
conjectures are also considered as strong (fine-grained) versions of 
the conjecture, ${\rm coC}_={\rm P}
\not\subseteq{\rm NP}$, which is believed because
${\rm coC}_={\rm P}\subseteq{\rm NP}$ leads to the collapse of the
polynomial-time hierarchy to the second level~\cite{Kobayashi,KobayashiICALP}.
Here, ${\rm coC}_={\rm P}$ is 
defined as follows.
\begin{definition}
A language $L$ is in ${\rm coC}_={\rm P}$ if and only if there exists
a non-deterministic polynomial-time Turing machine such that
if $x\in L$ then the number of accepting paths is not equal to
that of rejecting paths,
and
if $x\notin L$ then they are equal.
\end{definition}

These conjectures should be justified
by the following arguments.
First of all, it is true that
direct connections between our conjectures 
and SETH~\cite{IP01,IPZ01} 
(or NSETH~\cite{CGIMPS}) 
are not clear, because acceptance
criteria are different. (Our conjectures are based on
gap functions, while SETH 
and NSETH
are on $\#$P functions.)
However, at this moment, the only known way 
of deciding $gap(f)\neq0$ or $gap(f)=0$ is
to solve $\#$SAT problems. Even if
$f$ is restricted to $k$-CNF, the current fastest algorithm~\cite{CW}
to solve $\#$SAT satisfies $a\to0$ as $k\to\infty$,
and therefore it is true for more general circuits such as NC circuits or
general polynomial-size Boolean circuits.
Furthermore, although many non-deterministic algorithms
to decide whether $\#f=0$ or not
have been developed, NSETH is not yet refuted.
Deciding $gap(f)\neq0$ or not seems to be a similar type of problem.
Moreover, as is shown in Sec.~\ref{sec:discussion:NSETH}, 
Conjecture~\ref{conjecture:general} is reduced to 
UNSETH (Unique NSETH), which is a variant of NSETH
where $\#f=1$ is promised for the no case.

Why have we introduced three conjectures?
Let us explain relations among them.
Conjecture~\ref{conjecture:general}
considers the most general case, i.e., polynomial-size Boolean circuits,
and therefore the most ``stable" conjecture.
It would be safest to use that conjecture.
However, as we will see later, fine-grained quantum supremacy
results based on Conjecture~\ref{conjecture:general} can exclude
only super-polynomial time classical simulations.
We therefore introduce two ``less stable" conjectures,
Conjecture~\ref{conjecture:NC} and
Conjecture~\ref{conjecture:reversible}.
Conjecture~\ref{conjecture:NC} or Conjecture~\ref{conjecture:reversible} 
implies Conjecture~\ref{conjecture:general}, but no relation is known between
Conjecture~\ref{conjecture:NC} and Conjecture~\ref{conjecture:reversible}.
As is shown in Lemma~\ref{lemma:count} below,
Conjecture~\ref{conjecture:reversible}
is derived from an analogy of SETH.
Conjecture~\ref{conjecture:NC} is an analogy of
NC-SETH~\cite{AHVWW16}.

\begin{lemma}
\label{lemma:count}
For any $k$-CNF $f:\{0,1\}^n\to\{0,1\}$ with $k\in o(n)$, 
there exists a classical reversible circuit
$C:\{0,1\}^{n+\xi}\to\{0,1\}^{n+\xi}$ that uses only TOFFOLI and
NOT such that
$\xi\in o(n)$,
and
$C_{n+\xi}(x0^\xi)=f(x)$ for any $x\in\{0,1\}^n$.
\end{lemma}

We also introduce the following conjecture.

\begin{conjecture}
\label{conjecture:ETHN}
Any non-deterministic 
algorithm that decides whether $gap(f)\neq0$ or $gap(f)=0$
for
given 
(a description of) a 3-CNF with $m$ clauses,
$f:\{0,1\}^n\to\{0,1\}$,
needs $2^{\Omega(m)}$ time.
Here
\begin{eqnarray}
gap(f)\equiv\sum_{x\in\{0,1\}^n}(-1)^{f(x)}.
\end{eqnarray}
\end{conjecture}
Conjecture~\ref{conjecture:ETHN} is 
a ${\rm coC}_={\rm P}\not\subseteq{\rm NP}$-version of
Conjecture~\ref{conjecture:ETH}. As we have said,
the only known way of deciding $gap(f)\neq0$ or $gap(f)=0$ is
to solve the $\#$SAT problems, and therefore 
Conjecture~\ref{conjecture:ETHN} could be suggested by
ETH.

\begin{table}[htb]
\begin{tabular}{|c|c|c|}\hline
&functions&fine-grained of\\\hline
1~(ETH)&3-CNF&{\rm NP}$\not\subseteq${\rm P}\\\hline
2~(SETH)&CNF&{\rm NP}$\not\subseteq${\rm P}\\\hline
3~(NSETH)&Boolean (CNF)&{\rm coNP}$\not\subseteq${\rm NP}\\\hline
4&Boolean&{\rm coC}$_=${\rm P}$\not\subseteq${\rm NP}\\\hline
5&log-depth Boolean&{\rm coC}$_=${\rm P}$\not\subseteq${\rm NP}\\\hline
6&Reversible circuit&{\rm coC}$_=${\rm P}$\not\subseteq${\rm NP}\\\hline
7&3-CNF&{\rm coC}$_=${\rm P}$\not\subseteq${\rm NP}\\\hline
\end{tabular}
\label{table}
\caption{Summary of conjectures.}
\end{table}

\section{DQC1 model}
\label{sec:DQC1}
We first focus on the DQC1 model, 
which is defined as follows.
\begin{definition}
The $N$-qubit DQC1 model with a unitary $U$ is the following
quantum computing model.
\begin{itemize}
\item[1.]
The initial state is 
$
|0\rangle\langle0|\otimes \frac{I^{\otimes N-1}}{2^{N-1}},
$
where $I\equiv|0\rangle\langle0|+|1\rangle\langle1|$
is the two-dimensional identity operator.
\item[2.]
The unitary operator $U$ is applied on 
the initial state to generate
the state
$
U(
|0\rangle\langle0|\otimes \frac{I^{\otimes N-1}}{2^{N-1}})U^\dagger.
$
\item[3.]
The first qubit is measured in the computational basis.
If the output is 0 (1), then accept (reject).
\end{itemize}
\end{definition}
Note that throughout this paper
we consider only $poly(N)$-size $U$ without explicitly mentioning
it. 
The DQC1 model was originally introduced by Knill and Laflamme
to model NMR quantum computing~\cite{KL}, 
and since then many results have been
obtained on the model~\cite{KL,Poulin1,Poulin2,ShorJordan,
Passante,JordanWocjan,JordanAlagic,MFF,M,Kobayashi,nonclean,
KobayashiICALP}.
For example, the DQC1 model can solve several problems
whose classical efficient solutions are not known, such
as the spectral density estimation~\cite{KL}, 
testing integrability~\cite{Poulin1},
calculations of the fidelity decay~\cite{Poulin2}, 
and approximations of Jones and HOMFLY 
polynomials~\cite{ShorJordan,Passante,JordanWocjan}.
The acceptance probability $p_{acc}$ of the $N$-qubit
DQC1 model with a unitary $U$ is
\begin{eqnarray}
p_{acc}\equiv{\rm Tr}\Big[
(|0\rangle\langle0|\otimes I^{\otimes N-1})
U\Big(|0\rangle\langle0|\otimes\frac{I^{\otimes N-1}}{2^{N-1}}\Big)
U^\dagger\Big].
\end{eqnarray}
It is known that if $p_{acc}$ is
classically sampled in polynomial-time within
a multiplicative error $\epsilon<1$,
then the polynomial-time hierarchy 
collapses to the second level~\cite{Kobayashi,KobayashiICALP}.

Our results for the DQC1 model are the following three theorems.

\begin{theorem}
\label{theorem:general:DQC1}
Assume that Conjecture~\ref{conjecture:general} is true. Then
for any constant $a>0$ and for infinitely many $n$,
there exists an $N$-qubit DQC1 model, where $N=poly(n)$,
whose acceptance probability
$p_{acc}$ cannot be classically sampled
in time $2^{(1-a)n}$ within a multiplicative
error $\epsilon<1$.
\end{theorem}

\begin{theorem}
\label{theorem:NC:DQC1}
Assume that Conjecture~\ref{conjecture:NC} is true. Then
for any constant $a>0$ and for infinitely many $N$,
there exists an $N$-qubit DQC1 model
whose acceptance probability
$p_{acc}$
cannot be classically sampled
in time $2^{(1-a)(N-3)}$ within a multiplicative
error $\epsilon<1$.
\end{theorem}

\begin{theorem}
\label{theorem:reversible:DQC1}
Assume that Conjecture~\ref{conjecture:reversible} is true.
Then
for any constant $a>0$ and for infinitely many $N$,
there exists an $N$-qubit DQC1 model
whose acceptance probability
$p_{acc}$
cannot be classically sampled
in time $2^{(1-a)(N-\xi-2)}$ within a multiplicative
error $\epsilon<1$.
\end{theorem}

Theorem~\ref{theorem:general:DQC1} excludes only super-polynomial-time
classical simulations, because
although we know $N=poly(n)$,
we do not know the degree of the polynomial.
For example, $N$ might be $N=n^{10}$, 
and in this case,
the theorem says that the $N$-qubit DQC1 model 
cannot be classically
sampled in $2^{(1-a)N^{\frac{1}{10}}}$-time, which is 
superpolynomial (sub-exponential)
but not exponential.
The other two theorems exclude exponential-time simulations,
because $\xi\in o(n)$ and therefore
$2^{(1-a)(N-\xi-2)}=2^{(1-a)N+o(N)}$.
Because no relation is known between 
Conjecture~\ref{conjecture:NC}
and Conjecture~\ref{conjecture:reversible},
these two theorems are viewed as parallel results.

\section{HC1Q model}
\label{sec:HC1Q}
We next show similar fine-grained quantum supremacy results for 
another sub-universal quantum computing model, namely,
the Hadamard-classical circuit with one-qubit (HC1Q) model,
which is defined as follows.
\begin{definition}
The $N$-qubit HC1Q model with a classical reversible
circuit $C:\{0,1\}^N\to\{0,1\}^N$ is the following
quantum computing model.
\begin{itemize}
\item[1.]
The initial state is $|0^N\rangle$,
where $|0^N\rangle$ means $|0\rangle^{\otimes N}$.
\item[2.]
The operation 
$
(H^{\otimes N-1}\otimes I)
C(H^{\otimes N-1}\otimes I)
$
is applied on the
initial state. Here, 
$
H\equiv\frac{1}{\sqrt{2}}(|0\rangle+|1\rangle)\langle0|
+\frac{1}{\sqrt{2}}(|0\rangle-|1\rangle)\langle1|
$
is the Hadamard gate, and
the classical circuit $C$ is applied ``coherently".
\item[3.]
Some qubits are measured in the computational basis.
\end{itemize}
\end{definition}
Note that throughout this paper
we consider only $poly(N)$-size $C$ without explicitly mentioning
it. 
The HC1Q model is in the second level of the Fourier hierarchy~\cite{FH2}
where several useful circuits, such as those for
Shor's factoring algorithm~\cite{Shor}
and Simon's algorithm~\cite{Simon}, are
placed.
Furthermore, it was shown in Ref.~\cite{HC1Q} that
output probability distributions of the HC1Q model cannot be
classically sampled in polynomial-time within a multiplicative
error $\epsilon<1$ unless the polynomial-time hierarchy collapses
to the second level.

Our results are the following two theorems.

\begin{theorem}
\label{theorem:general:HC1Q}
Assume that Conjecture~\ref{conjecture:general} is true. Then
for any constant $a>0$ and for infinitely many $n$,
there exists an $N$-qubit HC1Q model, where $N=poly(n)$,
whose acceptance probability
$p_{acc}$
cannot be classically sampled
in time $2^{(1-a)n}$ within a multiplicative
error $\epsilon<1$.
\end{theorem}

\begin{theorem}
\label{theorem:reversible:HC1Q}
Assume that Conjecture~\ref{conjecture:reversible} is true.
Then
for any constant $a>0$ and for infinitely many $N$,
there exists an $N$-qubit HC1Q model
whose acceptance probability
$p_{acc}$
cannot be classically sampled
in time $2^{(1-a)(N-\xi-2)}$ within a multiplicative
error $\epsilon<1$.
\end{theorem}

The first theorem excludes super-polynomial-time classical simulations,
while the second one does exponential-time ones.

\section{Universal models}
\label{sec:univ}
\subsection{Clifford and $T$}
We also study universal quantum computing with Clifford and $T$ gates,
where 
$
T\equiv |0\rangle\langle0|+e^{i\pi/4}|1\rangle\langle1|,
$
and Clifford gates are $H$, 
$S\equiv|0\rangle\langle0|+i|1\rangle\langle1|$, and
$CZ\equiv I^{\otimes 2}-2|11\rangle\langle11|$.
For an $N$-qubit 
quantum circuit $V$ that consists of only Clifford and $T$ gates,
we define the acceptance probability $p_{acc}$ by
\begin{eqnarray}
p_{acc}\equiv|\langle0^N|V|0^N\rangle|^2.
\end{eqnarray}
Due to the Gottesman-Knill theorem~\cite{GK}, $p_{acc}$ can be exactly
calculated in $poly(N)$ time if
$V$ contains at most $O(\log(N))$ number of $T$ gates.
The brute-force classical calculation of $p_{acc}$ needs time that
scales in an exponential of $t$, where $t$ is the number of $T$ gates.

For Clifford-$T$ quantum computing, we can show the following two
theorems. (Theorem~\ref{theorem:ETH} was also shown
in Ref.~\cite{Huang2} independently.)

\begin{theorem}
\label{theorem:ETH}
Assume that Conjecture~\ref{conjecture:ETH} is true.
Then for infinitely many $t$ there exists
Clifford-$T$ quantum circuit with $t$ $T$ gates
whose acceptance probability $p_{acc}$
cannot be calculated in $2^{o(t)}$ time
within an additive error smaller than $2^{-\frac{3t+14}{7}}$.
\end{theorem}

\begin{theorem}
\label{theorem:ETHN}
Assume that Conjecture~\ref{conjecture:ETHN} is true.
Then for infinitely many $t$ there exists
a Clifford-$T$ quantum circuit with $t$ $T$ gates
whose acceptance probability $p_{acc}$ cannot be
classically sampled
in $2^{o(t)}$ time
within a multiplicative error $\epsilon<1$.
\end{theorem}

Note that Theorem~\ref{theorem:ETH} considers 
the calculation (not sampling) of output probability distributions 
(i.e., strong simulation).
Strong simulation
is 
$\#$P-hard, which means that it is hard even for
quantum computing. However, it is still meaningful to study
strong simulations of quantum computing because several quantum
computing models, such as Clifford circuits and 
match-gate circuits~\cite{match},
are known to be possible to strongly simulate.

There is a classical algorithm that computes $p_{acc}$ within a 
constant multiplicative error 
and in $\sim2^{\beta t}$ time with a non-trivial factor
$\beta\simeq 0.47$~\cite{BravyiGosset}.
Our second result (Theorem~\ref{theorem:ETHN}) 
therefore suggests that improving 
$2^{\beta t}$ to some sub-exponential time,
say $2^{\sqrt{t}}$,
is impossible (under the fine-grained conjecture).
Note that Ref.~\cite{DalzellPhD} showed that $\beta$ must be
$\beta>1/135\simeq0.0074$ under the conjecture,
\begin{eqnarray}
\mbox{MAJ-ZEROS}\not\in
\Sigma_3{\rm TIME}\left(\frac{2^{n/5}}{92}-\frac{n}{8}\right),
\end{eqnarray}
where $\Sigma_3$ is the third level of the polynomial-time hierarchy,
and MAJ-ZEROS is the problem of deciding whether $gap(f)>0$ or not
given a degree-3 polynomial $f$ in $n$ variables over the field
${\mathbb F}_2$.

The Pauli-based computation (PBC)~\cite{BravyiSmithSmolin} 
is a universal quantum
computing model closely related to the Clifford$+T$ quantum computing,
which is defined as follows.
\begin{definition}
The $t$-qubit Pauli-based quantum computation (PBC) is the following quantum
computing model.
\begin{itemize}
\item[1.]
The initial state is $|H\rangle^{\otimes t}$,
where
$
|H\rangle\equiv\cos\frac{\pi}{8}|0\rangle+\sin\frac{\pi}{8}|1\rangle
$
is a magic state.
\item[2.]
Non-destructive Pauli measurements are done adaptively.
(Here, adaptively means that a measurement basis depends on previous
measurement results.)
\item[3.]
The measurement results are finally classically processed.
\end{itemize}
\end{definition}

For the PBC, we show the following result.
\begin{theorem}
\label{theorem:PBC}
Assume that Conjecture~\ref{conjecture:ETHN} is true.
Then, for infinitely many $t$ there exists 
a $t$-qubit PBC whose output probability distributions
cannot be classically sampled in $2^{o(t)}$ time
within a multiplicative error $\epsilon<1$.
\end{theorem}

\subsection{$H$ and classical gates}
When we consider quantum computing over Clifford and $T$ gates,
we are interested in the number $t$ of $T$ gates, because
$T$ gates are ``quantum resources".
In a similar way, when we consider quantum computing over
classical gates and $H$, we are interested in the number $h$ of $H$ gates.
Since the number $h$ of $H$ in the $N$-qubit HC1Q model is
$h=2(N-1)$, Theorem~\ref{theorem:reversible:HC1Q} leads to the following
corollary.
\begin{corollary}
Assume that Conjecture~\ref{conjecture:reversible}
is true. Then for any constant $a>0$ and for
infinitely many $h$, there exists a quantum circuit with
classical gates and $h$ $H$ gates whose output probability 
distributions
cannot be classically sampled in time $2^{(1-a)\frac{h}{2}+o(h)}$
within a multiplicative error $\epsilon<1$.
\end{corollary}
Actually, we can show the same result with the more stable
conjecture,
Conjecture~\ref{conjecture:general}:
\begin{theorem}
\label{theorem:H}
Assume that Conjecture~\ref{conjecture:general}
is true. Then for any constant $a>0$ and for
infinitely many $h$, there exists a quantum circuit with
classical gates and $h$ $H$ gates whose output probability 
distributions
cannot be classically sampled in time 
$2^{(1-a)(\frac{h}{2}-\frac{1}{2})}$
within a multiplicative error $\epsilon<1$.
\end{theorem}

For strong simulation, we can show the following.
\begin{theorem}
\label{theorem:strongH}
Assume that Conjecture~\ref{conjecture:SETH} is true. Then, for any constant $a>0$ and for
infinitely many $h$, there exists an $N$-qubit quantum circuit 
$V$ over classical gates and $h$ $H$ gates such that
$N=poly(h)$ and
the classical exact calculation of 
$\|(I^{\otimes N-1}\otimes |1\rangle\langle1|)V|0^N\rangle\|^2$
cannot be done in deterministic $2^{(1-a)h}$ time. 
\end{theorem}

\subsection{$H$, $T$, and $CZ$ gates}
Strong simulation of the IQP (with $Z$, $CZ$, and $CCZ$) 
is possible for $a<0.0035$
by reducing the problem to counting the number of solutions
of polynomials~\cite{Tamaki}.
Here, the IQP model
is defined as follows~\cite{IQP1}.
\begin{definition}
The $N$-qubit IQP model with a unitary $D$ is the following quantum
computing model, where $D$ consists of only $Z$-diagonal gates
(such as $Z$, $CZ$, $CCZ$, and $e^{i\theta Z}$, etc.).
\begin{itemize}
\item[1.]
The initial state is $|0^N\rangle$.
\item[2.]
The operation $H^{\otimes N}DH^{\otimes N}$ is applied
on the initial state.
\item[3.]
All qubits are measured in the computational basis.
\end{itemize}
\end{definition}

By using a similar technique, we can show the following.

\begin{theorem}
\label{theorem:tamaki}
Let $U$ be an $N$-qubit $poly(N)$-size quantum circuit over $H$, $T$,
and $CZ$. For any $a,b\in\{0,1\}^N$, $\langle a|U|b\rangle$ can be
exactly calculated in deterministic $2^{0.984965h+o(h)}$ time,
where $h$ is the number of $H$ gates in $U$.
\end{theorem}

This is a new non-trivial classical simulation of quantum circuits
over $H$, $T$, and $CZ$. 
The technique of Ref.~\cite{Tamaki} can compute the number of roots
of polynomials over ${\mathbb F}_q$, but it cannot be used for those over
${\mathbb Z}_m$. The proof of Theorem~\ref{theorem:tamaki}
considers some ${\mathbb Z}_m$. This technique itself seems to be of use.

\section{Stabilizer rank}
\label{sec:stabilizer}
We can also show lowerbounds of the stabilizer 
rank~\cite{BravyiSmithSmolin}, which is defined as follows.
\begin{definition}[Stabilizer rank]
\label{definition:stabilizer_rank}
The stabilizer rank $\chi(|\psi\rangle)$ of an $n$-qubit pure state
$|\psi\rangle$ is the smallest integer $k$ such that $|\psi\rangle$
can be written as
\begin{eqnarray}
|\psi\rangle=\sum_{j=1}^k c_j C_j|0^n\rangle,
\label{decomposition}
\end{eqnarray}
where each $c_j$ is a coefficient 
and each $C_j$ is a Clifford circuit.
\end{definition}

Note that the original definition of the stabilizer rank 
(Definition~\ref{definition:stabilizer_rank})
does not care about
computational complexity of $\{c_j\}_j$ and $\{C_j\}_j$: 
the minimum of $k$ is taken over all decompositions of $|\psi\rangle$
in the form of Eq.~(\ref{decomposition}).
In this paper, however, we consider only decompositions in the 
form of Eq.~(\ref{decomposition}) such that
there exists
a $poly(n)$-time classical deterministic algorithm that, on input $j$,
outputs $c_j$ and a classical description of
$C_j$. Such an additional restriction
is relevant when we study the stabilizer rank in the context of
classical simulations of quantum computing.

The stabilizer rank is directly connected to the time complexity
of classical simulations of
quantum computing. For example,
by using the well-known gadget 
\begin{eqnarray*}
(I\otimes|0\rangle\langle0|)
(I\otimes H)CZ(I\otimes H)
(|\psi\rangle\otimes|A\rangle)=\frac{1}{\sqrt{2}}T|\psi\rangle,
\end{eqnarray*}
where
$
|A\rangle\equiv\frac{1}{\sqrt{2}}(|0\rangle+e^{i\frac{\pi}{4}}|1\rangle)
$
is a magic state,
we can easily show
that for any universal quantum
circuit $U$ that uses Clifford gates and $t$ $T$ gates, 
there exists a Clifford circuit $V$ such that
\begin{eqnarray}
\langle0^n|U|0^n\rangle=
\sqrt{2^t}\langle0^{n+t}|V(|0^n\rangle\otimes|A\rangle^{\otimes t}).
\end{eqnarray}
Since
\begin{eqnarray}
&&\langle0^{n+t}|V(|0^n\rangle\otimes|A\rangle^{\otimes t})\nonumber\\
&=&\sum_{j=1}^{\chi(|A\rangle^{\otimes t})}
c_j\langle0^{n+t}|V(I^{\otimes n}\otimes C_j)|0^{n+t}\rangle,
\end{eqnarray}
and each
$\langle0^{n+t}|V(I^{\otimes n}\otimes C_j)|0^{n+t}\rangle$
can be computed in $poly(n+t)=poly(n)$ time,
the value $\langle 0^n|U|0^n\rangle$ can be calculated
in $\chi(|A\rangle^{\otimes t})poly(n)$ time
(assuming that there exists a classical $poly(n)$-time
algorithm that, on input $j$, outputs $c_j$ and
classical description of $C_j$).
In this way, the stabilizer rank is directly connected
to the time complexity of classical simulations.
We do not know how to calculate the
exact value of the stabilizer rank,
and therefore finding better upperbounds of the stabilizer rank is
essential.
Several non-trivial upperbounds are known ~\cite{BravyiSmithSmolin},
such as
$\chi(|A\rangle^{\otimes 6})\le7$, which means
\begin{eqnarray}
\chi(|A\rangle^{\otimes t})\le2^{t\frac{\log_27}{6}}\simeq 2^{0.468t}.
\end{eqnarray}
It is open how much we can improve this upperbound.
If we believe ${\rm BQP}\neq{\rm BPP}$, it is clear that
$\chi(|A\rangle^{\otimes t})\le poly(t)$ is impossible.
It was conjectured in Ref.~\cite{BravyiSmithSmolin} that
$\chi(|A^{\otimes t}\rangle)\ge2^{\Omega(t)}$.
Only known lowerbound is the
very weak one~\cite{BravyiSmithSmolin}
\begin{eqnarray}
\chi(|A^{\otimes t}\rangle)\ge\Omega(t^{\frac{1}{2}}),
\end{eqnarray}
which is not enough to show the conjecture.

Based on Conjecture~\ref{conjecture:ETH} (ETH), we can show the following.
\begin{theorem}
\label{theorem:stabilizer_rank}
Assume that Conjecture~\ref{conjecture:ETH} is true. 
Let $|\Psi\rangle$ be
a resource of $t$ TOFFOLI gates.
Then,
$\chi(|\Psi\rangle)\ge2^{\Omega(t)}$.
\end{theorem}

Here, the meaning of the statement
``$|\Psi\rangle$ is a resource of $t$ TOFFOLI gates"
is defined as follows.
\begin{definition}
\label{definition:resource}
Let $g$ be a non-Clifford gate (such as $T$, $CCZ$, or TOFFOLI).
We say that an $r$-qubit state
$|\Psi\rangle$ is a resource of $t$ $g$ gates if the following three
conditions are all satisfied.
\begin{itemize}
\item[1.]
$r=poly(t)$.
\item[2.]
For any $n$-qubit quantum circuit $U$ over Clifford gates and
$t$ $g$ gates, there exists an $(n+r)$-qubit Clifford circuit
$V$ such that
\begin{eqnarray}
\frac{
(I^{\otimes n}\otimes |0\rangle\langle0|^{\otimes r})
V(|0^n\rangle\otimes|\Psi\rangle)
}
{
\|(I^{\otimes n}\otimes |0\rangle\langle0|^{\otimes r})
V(|0^n\rangle\otimes|\Psi\rangle)\|}
=U|0^n\rangle.
\end{eqnarray}
\item[3.]
The quantity
\begin{eqnarray}
\|(I^{\otimes n}\otimes |0\rangle\langle0|^{\otimes r})
V(|0^n\rangle\otimes|\Psi\rangle)\|
\end{eqnarray}
is computable in $poly(r)$ time. 
\end{itemize}
\end{definition}
For example, it is easy to verify that
$|A\rangle^{\otimes t}$ is a resource of $t$ $T$ gates.

In particular, if we take
\begin{eqnarray}
|\Psi\rangle&=&|CCZ\rangle^{\otimes t},\\
|\Psi\rangle&=&|A\rangle^{\otimes 7t},
\end{eqnarray}
where $|CCZ\rangle$ is the resource of
a single $CCZ$ gate,
in Theorem~\ref{theorem:stabilizer_rank},
we obtain the following corollary.

\begin{corollary}
\begin{eqnarray}
\chi(|CCZ\rangle^{\otimes t})\ge 2^{\Omega(t)},\\
\chi(|A\rangle^{\otimes t})\ge 2^{\Omega(t)}.
\end{eqnarray}
\end{corollary}

\section{Discussion}
\label{sec:discussion}

\subsection{Optimality}

The acceptance probability of an $N$-qubit DQC1 model
can be exactly sampled in $2^{N+o(N)}$ time.
(Generate a uniformly random $(N-1)$-bit string $a$, and accept with
the probability
$\langle 0a|U^\dagger(|0\rangle\langle0|\otimes I^{\otimes N-1})U|0a\rangle$.)
Hence Theorem~\ref{theorem:NC:DQC1} and Theorem~\ref{theorem:reversible:DQC1}
are optimal.

The acceptance
probability of any $N$-qubit HC1Q model can be calculated in
$2^{N+\log(poly(N))-1}$ time. 
Theorem~\ref{theorem:reversible:HC1Q} is therefore optimal.
In fact, by a straightforward calculation, 
\begin{eqnarray}
&&\langle z|
(H^{\otimes N-1}\otimes I)
C
(H^{\otimes N-1}\otimes I)
|0^N\rangle\nonumber\\
&=&
\frac{1}{2^{N-1}}\sum_{x\in\{0,1\}^{N-1}}
(-1)^{\sum_{j=1}^{N-1}z_jC_j(x0)}
\delta_{z_N,C_N(x0)}
\end{eqnarray}
for any reversible circuit
$C:\{0,1\}^N\to\{0,1\}^N$ and $z\in\{0,1\}^N$,  
where $C_j(x0)\in\{0,1\}$ is the $j$th bit of $C(x0)\in\{0,1\}^N$.
If $C$ is $poly(N)$ size,
each term of the exponential sum can be computed
in $poly(N)$ time, and to sum all of them needs
$2^{N-1}$ time. The total time is therefore
\begin{eqnarray}
poly(N)\times 2^{N-1}=2^{N+\log(poly(N))-1}.
\end{eqnarray}

\subsection{Fine-grained supremacy for the IQP model}
\label{sec:IQP}
Ref.~\cite{Dalzell}
showed a fine-grained result on
the hardness of classically sampling output probability
distributions of the IQP model within a multiplicative error. 
To show fine-grained quantum supremacy of the IQP model,
they introduced a
conjecture,
so-called poly3-NSETH$(a)$, which
is the same as Conjecture~\ref{conjecture:general}
except that $f$ is restricted to be 
polynomials over the field $\mathbb{F}_2$
with degree at most 3. 
A reason why
$f$ is restricted to be polynomials is that IQP circuits
can calculate polynomials over $\mathbb{F}_2$ without introducing
any ancilla qubit. (For the QAOA model, $n$ ancilla qubits are
necessary~\cite{Dalzell}.) 
However, a disadvantage of poly3-NSETH$(a)$ is that
it is violated when $a<0.0035$~\cite{Tamaki}.
(It was argued in Ref.~\cite{Dalzell} that improving 
the algorithm of Ref.~\cite{Tamaki} will not rule out 
poly3-NSETH$(a)$ with $a\ge0.5$,
and therefore they conjecture poly3-NSETH$(a)$ for
$a\ge0.5$.) 
Our conjectures, on the other hand, consider
general Boolean circuits, which are more stable than those
on degree-3 polynomials.
The reason why it is possible is that there is no gate
restriction for the DQC1 and HC1Q models.
Since general Boolean circuits cannot be efficiently represented by
systems of equations of polynomials, the technique of Ref.~\cite{Tamaki}
cannot be used to refute
our conjectures
on general Boolean circuits.

\subsection{Fine-grained supremacy of the Boson sampling model}
\label{sec:BS}
Ref.~\cite{Dalzell} also studied fine-grained quantum supremacy
of the Boson Sampling model. They introduced a conjecture, so-called
per-int-NSETH$(b)$, which states that deciding
whether the permanent of a given $n\times n$ integer matrix
is nonzero needs non-deterministic $2^{bn}$ time.
In this case, no value of $b<1$ is ruled out by
known algorithms~\cite{Dalzell}.
It is not clear how per-int-NSETH$(b)$ and our conjectures are
related with each other. At least we can show by
using Ryser's formula and Chinese remainder theorem that
if $\#f$ of $n$-variable Boolean circuits are calculated in
time $2^{(1-a)n}$, then permanents of $n\times n$ integer
matrices are calculated in $2^{(1-a)n}$ time.

\subsection{Restricting to CNF}
In Conjecture~\ref{conjecture:general}, we have assumed that $f$ is any
polynomial-size Boolean circuit. It is still reasonable
to consider Conjecture~\ref{conjecture:general}
with restricting $f$ to be $k$-CNF formulas
while keeping the $a>0$ condition.
(In fact, at this moment, the only known way
of deciding whether $gap(f)\neq0$ or $gap(f)=0$ is to solve
$\#{\rm SAT}$ problems. The current fastest algorithm~\cite{CW}
to solve $\#{\rm SAT}$ of $k$-CNF does not contradict
the condition $a>0$.)
In this case, required quantum circuits should be
simpler than those for general polynomial-size Boolean circuits.

\subsection{NSETH}
\label{sec:discussion:NSETH}
Conjectures of the present paper and
poly3-NSETH$(a)$ of Ref.~\cite{Dalzell}
are fine-grained versions of 
${\rm coC}_={\rm P}\not\subseteq {\rm NP}$.
It is interesting to ask whether
we can use NSETH~\cite{CGIMPS}, which is a fine-grained version
of ${\rm coNP}\not\subseteq{\rm NP}$, to show fine-grained
quantum supremacy.

At this moment, we do not know whether we can show
any fine-grained quantum supremacy result under NSETH.
At least, we can show that proofs of 
our theorems
(and those of Ref.~\cite{Dalzell})
cannot be directly applied to the case of NSETH.
To see it,
let us consider the following ``proof".
(For details, see Sec.~\ref{sec:proofs}.)
Given a Boolean circuit $f:\{0,1\}^n\to\{0,1\}$,
we first construct an $m\equiv n+\xi$
qubit quantum circuit $V$ such that
$0<\eta<1$ if $\# f=0$, and
$\eta=0$ if $\# f>0$,
where $\xi=poly(n)$ and
$\eta\equiv|\langle0^m|V|0^m\rangle|^2$.
By using Lemma~\ref{lemma},
we next construct the $N\equiv m+2=n+\xi+2$
qubit DQC1 model  
whose acceptance probability is
\begin{eqnarray}
p_{acc}=\frac{4\eta(1-\eta)}{2^m}.
\end{eqnarray}
Then, if we assume that $p_{acc}$ is classically sampled
within a multiplicative error $\epsilon<1$ and in 
$2^{(1-a)n}$ time,
then NSETH is violated.

This ``proof" seems to work, but
actually we do not know how to construct
such $V$. In fact, the following lemma suggests
that we cannot construct such $V$.

\begin{lemma}
\label{lemma:noV}
Assume that given a Boolean circuit $f:\{0,1\}^n\to\{0,1\}$,
we can construct an $m\equiv n+\xi$
qubit quantum circuit $V$ such that
$0<\eta<1$ if $\# f=0$, and
$\eta=0$ if $\# f>0$,
where $\xi=poly(n)$ and
$\eta\equiv|\langle0^m|V|0^m\rangle|^2$.
Then ${\rm coNP}\subseteq{\rm coC}_={\rm P}$. 
\end{lemma}
However, there is an oracle
$A$ such that  
${\rm coNP}^A\not\subseteq{\rm coC}_={\rm P}^A$~\cite{Beigel},
which suggests that such a containment is unlikely.

We do not know whether our conjectures can be reduced to
more standard ones, such as SETH and NSETH.
At least, we can show that Conjecture~\ref{conjecture:general} 
is reduced to UNSETH (Unique NSETH) that is equal to
NSETH (Conjecture~\ref{conjecture:NSETH}) 
except that $\#f=1$ is promised for the no case.
This means that if UNSETH is true, then Conjecture~\ref{conjecture:general}
is also true.
In fact, for a given polynomial-size Boolean circuit
$f:\{0,1\}^n\to\{0,1\}$, define the polynomial-size Boolean
circuit $g:\{0,1\}^{n+1}\to\{0,1\}$ by
\begin{eqnarray}
g(x,x_{n+1})=[x_{n+1}\wedge \bar{f}(x)]\vee[\bar{x}_{n+1}\wedge
(\wedge_{j=1}^nx_j)]
\end{eqnarray}
for any $x\in\{0,1\}^n$. Then, 
\begin{eqnarray}
gap(g)&=&\sum_{x\in\{0,1\}^n}\sum_{x_{n+1}\in\{0,1\}}
(-1)^{g(x,x_{n+1})}\nonumber\\
&=&
2^n-2+
\sum_{x\in\{0,1\}^n}(-1)^{\bar{f}(x)},
\end{eqnarray}
and therefore 
if $\#f=0$ then $gap(g)\neq0$ and
if $\#f=1$ then $gap(g)=0$.

\subsection{Other conjectures}
In addition to SETH, NC-SETH, and NSETH, there exists
another conjecture, $\oplus$-SETH,
which asserts that for any $a>0$ there exists
a large integer $k$ such that $k$-CNF-$\oplus$SAT cannot be
computed in time $O(2^{(1-a)n})$~\cite{ParitySETH}. Here, 
$k$-CNF-$\oplus$SAT is the problem of computing the number of
satisfying assignments of a given $k$-CNF formula modulo two.
It is interesting to study 
whether we can find any fine-grained quantum supremacy
based on $\oplus$-SETH.
It is also open whether we can show 
any fine-grained quantum supremacy
under other conjectures that are not based on
SAT, such as 3-SUM~\cite{3-SUM}
and All-Pairs Shortest Paths problem (APSP)~\cite{APSP}.

\subsection{Additive error sampling}
In this paper we have considered multiplicative error sampling.
Multiplicative error sampling is a somehow strict notion of approximation
(it is almost equivalent to exact sampling),
and therefore more practical approximate sampling should be considered
for experimental realizations.

It is known that output probability distributions of
several sub-universal quantum computing models,
such as the Boson Sampling model~\cite{BS}, 
the IQP model~\cite{IQP2}, the random circuit
model~\cite{random}, and the DQC1 model~\cite{noteM,M}, 
cannot be classically sampled in
polynomial time within an additive error unless the polynomial-time
hierarchy collapses to the third level.
(Note that in addition to the infiniteness of the polynomial-time hierarchy,
we need another additional conjecture about the average-case
$\#$P-hardness of computing a certain function within a multiplicative 
error. Furthermore,
for the Boson Sampling model, another conjecture,
the anti-concentration conjecture, is also required.) 
Here, additive error sampling is defined as follows.
\begin{definition}
We say that a probability distribution $\{p_z\}_z$ is
classically sampled in time $T$ within an additive error $\epsilon$
if there exists a classical probabilistic 
algorithm that runs in time $T$ 
such that 
\begin{eqnarray}
\sum_z|p_z-q_z|\le\epsilon,
\end{eqnarray}
where $q_z$ is the probability that
the classical algorithm outputs $z$.
\end{definition}

It is an important open question whether any fine-grained version of
those additive-error results is possible or not.
The standard proof for the additive error supremacy is the combination
of the Stockmeyer's theorem and the average-case-hardness 
assumption~\cite{BS,IQP2}.
In this direction, two concrete open problems are 
(1)can we show the
``exponential-time variant" of the Stockmeyer's theorem? and
(2)can we show the average-case hardness for the fine-grained complexity
conjectures?
Regarding the second point,
some uniform average-case hardness results are known for fine-grained
complexity conjectures~\cite{FGaverage}, 
so we might be able to use them.

\section{Proofs}
\label{sec:proofs}
In this section, we provide proofs postponed.

\subsection{Proof of Lemma~\ref{lemma:count}}
\begin{figure}[htbp]
\begin{center}
\includegraphics[width=0.4\textwidth]{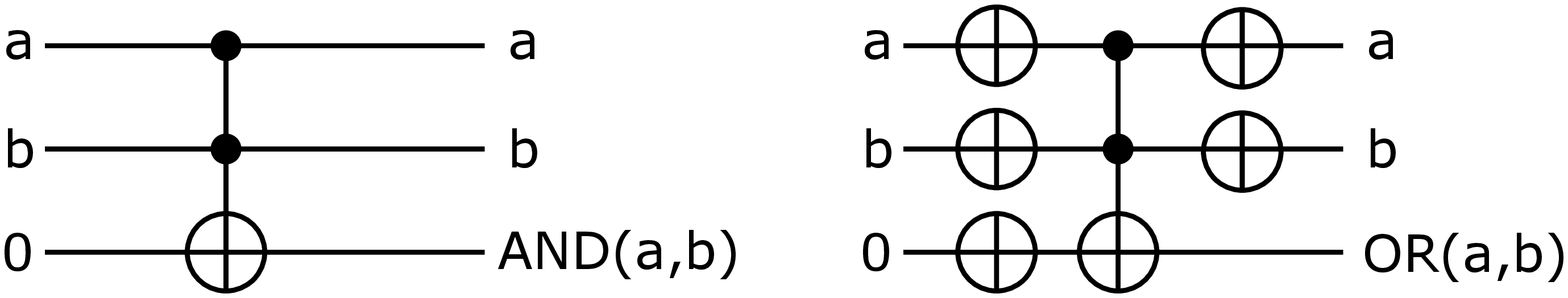}
\end{center}
\caption{Left: A simulation of an AND gate with a TOFFOLI gate.
Right: A simulation of an OR gate with a TOFFOLI gate and NOT gates.
}
\label{ANDOR}
\end{figure}

A $k$-CNF $f:\{0,1\}^n\to\{0,1\}$ consists of AND, OR, and NOT gates
that are defined by
\begin{eqnarray}
{\rm AND}(a,b)&=&ab,\\
{\rm OR}(a,b)&=&1-(1-a)(1-b),\\
{\rm NOT}(a)&=&a\oplus 1,
\end{eqnarray}
for any $a,b\in\{0,1\}$.
An AND gate can be simulated by
a TOFFOLI gate
by using a single ancilla bit initialized to 0 
(Fig.~\ref{ANDOR}, left). 
An OR gate can be simulated by
a TOFFOLI gate and NOT gates
by using a single ancilla bit initialized to 0 
(Fig.~\ref{ANDOR}, right).

\begin{figure}[htbp]
\begin{center}
\includegraphics[width=0.3\textwidth]{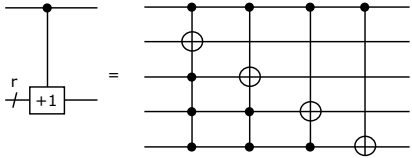}
\end{center}
\caption{
The counter operator $\Lambda_+^r$ with $r=4$.
}
\label{counter_example}
\end{figure}

Let us define the counter operator $\Lambda_+^r$ by
\begin{eqnarray}
\Lambda_+^r(|a\rangle\otimes|b\rangle)=|a\rangle\otimes
|b+a~({\rm mod}~2^r)\rangle,
\end{eqnarray}
where $a\in\{0,1\}$ and $b\in\{0,1,2,...,2^r-1\}$.
The counter operator $\Lambda_+^r$ can be constructed with
$r$ generalized TOFFOLI gates. 
For example, the construction for $r=4$ is given
in Fig.~\ref{counter_example}.
It is clear from the induction 
that $\Lambda_+^r$ for general $r$ is constructed
in a similar way.
Each generalized TOFFOLI gate can be decomposed as a linear number
of TOFFOLI gates with a single uninitialized
ancilla bit that
can be reused~\cite{Barenco}, and therefore a single $\Lambda_+^r$ requires
a single uninitialized ancilla bit.

\begin{figure}[htbp]
\begin{center}
\includegraphics[width=0.9\textwidth,angle=90]{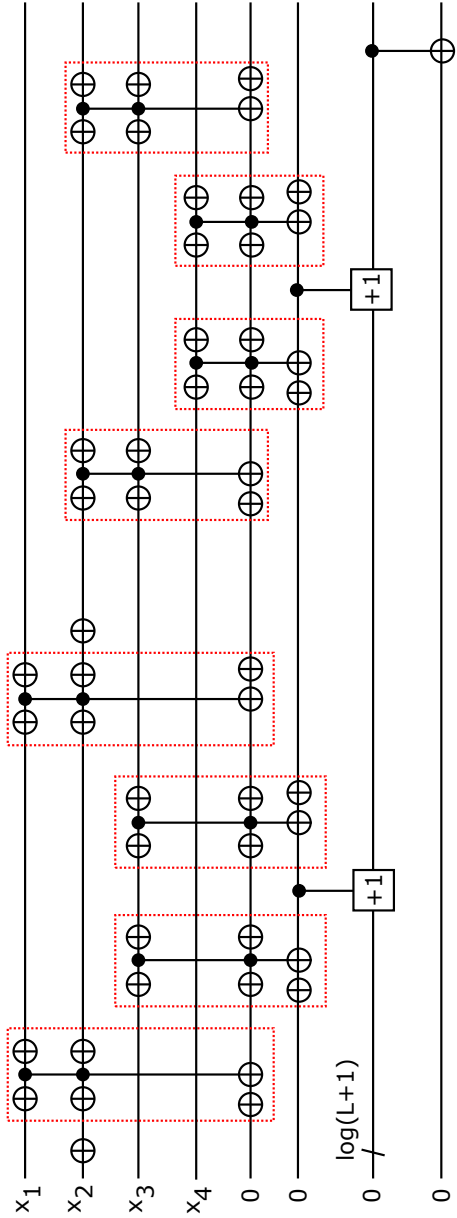}
\end{center}
\caption{
A circuit that evaluates
$(x_1\vee \bar{x}_2\vee x_3)\wedge(x_2\vee x_3\vee x_4)$.
Each circuit in a red dotted box is the reversible OR gate.
}
\label{counter}
\end{figure}

By using the counter operators, let us construct
the circuit of Fig~\ref{counter}, which computes
the $3$-CNF,
\begin{eqnarray}
(x_1\vee \bar{x}_2\vee x_3)\wedge(x_2\vee x_3\vee x_4).
\end{eqnarray}
For a $k$-CNF, $f:\{0,1\}^n\to\{0,1\}$, it is clear from the figure that 
\begin{itemize}
\item[1.]
$k-1$ ancilla bits initialized to 0
are necessary to calculate the value of
a single clause. However,
since these ancilla bits are reset to 0 after evaluating
a clause, these ancilla bits are reusable.
\item[2.]
To count the number of clauses that is 1,
$\log (L+1)$ ancilla bits are necessary, where $L$ is the number of
clauses. Note that
$\log (L+1)\in o(n)$, because
\begin{eqnarray}
\log (L+1)&\le& \log(L)+1\nonumber\\
&\le&\log \frac{2n(2n-1)...(2n-k+1)}{k!}+1\nonumber\\
&\le&\log \frac{(2n)^k}{k!}+1\nonumber\\
&\le& k+k\log n-k\log k+1.
\end{eqnarray}
\item[3.]
Each counter operator needs a single uninitialized ancilla bit.
Since the ancilla bit is reusable, only a single ancilla bit is enough 
throughout
the computation.
This ancilla bit can also be used for the final
$\log(L+1)$-bit TOFFOLI.
\item[4.]
Finally, a single ancilla bit that encodes $f(x)$
is necessary.
\end{itemize}
Hence, in total, the number $\xi$
of ancilla bits required is
\begin{eqnarray}
\xi=k-1+\log (L+1)+1+1=o(n).
\end{eqnarray}

\subsection{Proof of Theorem~\ref{theorem:general:DQC1}}
Let $f:\{0,1\}^n\to\{0,1\}$ be (a description of)
a polynomial-size Boolean circuit.
Let $\xi$ be the number of AND and OR gates in $f$. 
Since $f$ is polynomial-size, $\xi=poly(n)$.
Then by simulating each AND and OR in $f$ with TOFFOLI and NOT,
we can construct the $(n+\xi)$-qubit unitary operator $U$ 
that uses only $X$ and TOFFOLI such that 
\begin{eqnarray}
U(|x\rangle\otimes|0^\xi\rangle)
=
|junk(x)\rangle
\otimes
|f(x)\rangle
\end{eqnarray}
for
any $x\in\{0,1\}^n$, 
where
$junk(x)\in\{0,1\}^{n+\xi-1}$ is a certain bit string
whose detail is irrelevant here.
Define the $(n+\xi)$-qubit unitary operator $V$ by
\begin{eqnarray}
V\equiv 
(H^{\otimes n+\xi})
(I^{\otimes n+\xi-1}\otimes Z)
U
(H^{\otimes n}\otimes I^{\otimes \xi}),
\end{eqnarray}
where $H$ is the Hadamard gate and $Z$ is the Pauli $Z$ gate.
Then, 
\begin{eqnarray}
\eta&\equiv&|\langle 0^{n+\xi}|V|0^{n+\xi}\rangle|^2\nonumber\\
&=&|\langle +^{n+\xi}|(I^{\otimes n+\xi-1}\otimes Z)U
\frac{1}{\sqrt{2^n}}\sum_{x\in\{0,1\}^n}|x\rangle\otimes|0^\xi\rangle|^2
\nonumber\\
&=&|\langle +^{n+\xi}
|\frac{1}{\sqrt{2^n}}\sum_{x\in\{0,1\}^n}
(-1)^{f(x)}|junk(x)\rangle|f(x)\rangle|^2\nonumber\\
&=&\frac{gap(f)^2}{2^{2n+\xi}}.
\end{eqnarray}
(Note that the relation between the gap function and the DQC1 model was
also studied in Ref.~\cite{Datta}.)
If $gap(f)\neq 0$ then $0<\eta<1$.
If $gap(f)=0$ then $\eta=0$.
From Lemma~\ref{lemma} given below, by taking $m=n+\xi$,
we can construct the $N'\equiv n+\xi+1$ qubit DQC1 model
such that its acceptance probability 
is
\begin{eqnarray}
p_{acc}=\frac{4\eta(1-\eta)}{2^{n+\xi}}.
\end{eqnarray}
If $gap(f)\neq 0$ then $p_{acc}>0$.
If $gap(f)=0$ then $p_{acc}=0$.
An $m$-qubit TOFFOLI can be decomposed into a linear number of
TOFFOLI gates with
a single ancilla qubit~\cite{Barenco}.
In the construction, the ancilla qubit is not necessarily initialized,
and therefore the completely-mixed state $I/2$ can be used.
Hence the $N'$-qubit DQC1 model can be simulated
by the $N\equiv N'+1=n+\xi+2$ qubit DQC1 model.

Assume that there exists a
classical probabilistic algorithm that samples
$p_{acc}$ within a multiplicative error
$\epsilon<1$ 
and in time $2^{(1-a)n}$.
This means that
\begin{eqnarray}
|p_{acc}-q_{acc}|\le \epsilon p_{acc},
\end{eqnarray}
where $q_{acc}$ is the acceptance
probability of the $2^{(1-a)n}$-time classical probabilistic algorithm.
If $gap(f)\neq0$ then 
\begin{eqnarray}
q_{acc}\ge(1-\epsilon)p_{acc}>0,
\end{eqnarray}
and if $gap(f)=0$ then 
\begin{eqnarray}
q_{acc}\le(1+\epsilon)p_{acc}=0.
\end{eqnarray}
This means that there exists a non-deterministic
algorithm running
in time $2^{(1-a)n}$ such that
if $gap(f)\neq0$ then accepts and if $gap(f)=0$
then rejects.
However, it contradicts
to Conjecture~\ref{conjecture:general}.

\begin{figure}[htbp]
\begin{center}
\includegraphics[width=0.45\textwidth]{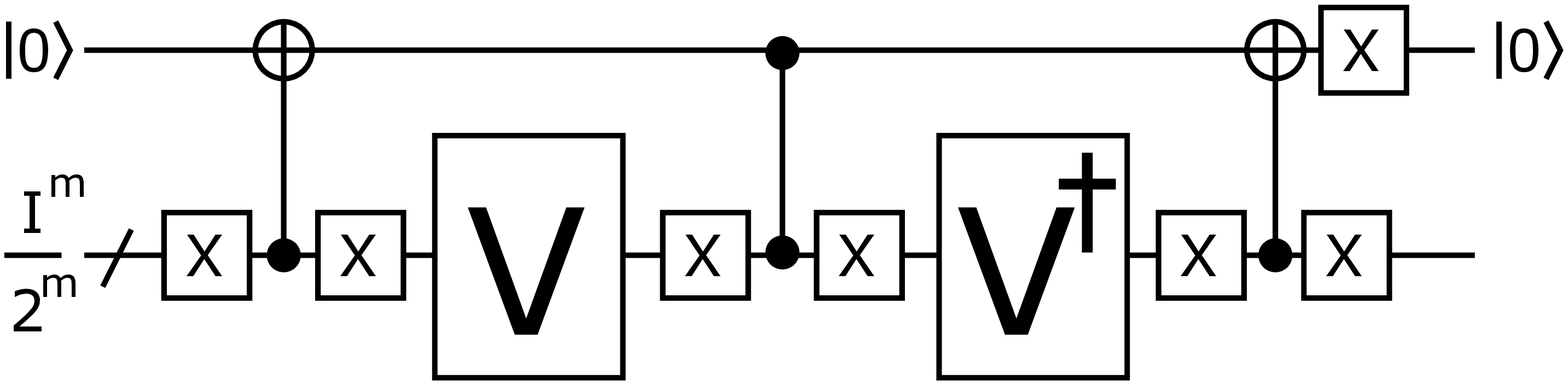}
\end{center}
\caption{The $(m+1)$-qubit DQC1 model constructed from an $m$-qubit
unitary $V$.
The line with the slash $/$ means the set of $m$ qubits.
A gate acting on this line is applied on each qubit.
The first qubit is measured in the computational basis.
If the measurement result is 0, as is indicated in the figure,
we accept. Otherwise, reject.
}
\label{DQC1_fig}
\end{figure}

\begin{lemma}\cite{Kobayashi}
\label{lemma}
Let $V$ be a quantum circuit acting on $m$ qubits.
From $V$, let us construct the $(m+1)$-qubit DQC1 circuit of 
Fig.~\ref{DQC1_fig}.
Then, the acceptance probability 
$p_{acc}$ 
of the DQC1 model (i.e., the probability of obtaining 0
in the computational-basis measurement), 
is
\begin{eqnarray}
p_{acc}=\frac{4\eta(1-\eta)}{2^m},
\end{eqnarray}
where 
$\eta\equiv|\langle0^m|V|0^m\rangle|^2$.
\end{lemma}
A proof of Lemma~\ref{lemma}
is obtained by a straightforward calculation.
(See also Ref.~\cite{Kobayashi}.)

\subsection{Proof of Theorem~\ref{theorem:general:HC1Q}}
\begin{figure}[htbp]
\begin{center}
\includegraphics[width=0.4\textwidth]{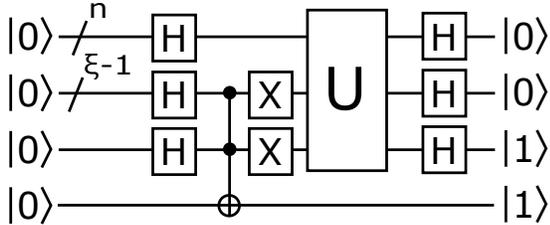}
\end{center}
\caption{The $(n+\xi+1)$-qubit HC1Q circuit used for the proof.
The first line with the slash $/$ means the set of $n$ qubits.
The second line with the slash $/$ means the set of $\xi-1$ qubits.
A gate acting on these lines is applied on each qubit.
All qubits are measured in the computational basis.
If the measurement result is $0^n0^{\xi-1}11$, as is indicated in the figure,
we accept.
Otherwise, reject.
}
\label{fig_HC1Q}
\end{figure}

Given (a description of) a Boolean circuit
$f:\{0,1\}^n\to\{0,1\}$, we again construct
the $(n+\xi)$-qubit unitary operator $U$ such that
\begin{eqnarray}
U(|x\rangle\otimes|0^\xi\rangle)=
|junk(x)\rangle
\otimes
|f(x)\rangle
\end{eqnarray}
for any $x\in\{0,1\}^n$, where $\xi=poly(n)$
and $junk(x)\in\{0,1\}^{n+\xi-1}$ is a bit string.
Note that $U$ uses only $X$ and TOFFOLI.
Consider the 
$N'\equiv n+\xi+1$ qubit HC1Q circuit in Fig.~\ref{fig_HC1Q}.
By a straightforward calculation,
the probability of obtaining the result $0^n0^{\xi-1}11$,
which we define as the acceptance probability $p_{acc}$,
is
\begin{eqnarray}
p_{acc}=\frac{gap(f)^2}{2^{2n+2\xi}}.
\end{eqnarray}
Then, if $gap(f)\neq0$, $p_{acc}>0$.
If $gap(f)=0$, $p_{acc}=0$. 
The $\xi$-qubit TOFFOLI used in the circuit of Fig.~\ref{fig_HC1Q}
can be decomposed into a linear number of TOFFOLI gates
with a single uninitialized ancilla qubit, which can be $|+\rangle$
state~\cite{Barenco}.
Therefore, the $N'$-qubit HC1Q model is simulated by
the $N\equiv N'+1=n+\xi+2$ qubit HC1Q model.

Assume that there exists a classical probabilistic algorithm
that samples $p_{acc}$ in time $2^{(1-a)n}$ and
within a multiplicative error $\epsilon<1$:
\begin{eqnarray}
|p_{acc}-q_{acc}|\le\epsilon p_{acc},
\end{eqnarray}
where $q_{acc}$ is the acceptance probability of the classical algorithm.
Then, if $gap(f)\neq0$,
\begin{eqnarray}
q_{acc}\ge(1-\epsilon)p_{acc}>0,
\end{eqnarray}
and
if $gap(f)=0$,
\begin{eqnarray}
q_{acc}\le(1+\epsilon)p_{acc}=0.
\end{eqnarray}
This means
that deciding $gap(f)\neq0$ or $gap(f)=0$ can be done
in non-deterministic $2^{(1-a)n}$ time, which contradicts
to Conjecture~\ref{conjecture:general}.

\subsection{Proof of Theorem~\ref{theorem:NC:DQC1}}
It is shown in Ref.~\cite{Cosentino} that any logarithmic depth Boolean circuit
$f:\{0,1\}^n\to\{0,1\}$ (that consists of AND, OR, and NOT) can
be implemented with a polynomial-size quantum circuit
$U$ acting on $n+1$ qubits such that
\begin{eqnarray}
U(|x\rangle\otimes|b\rangle)=
e^{ig(x)}
|x\rangle
\otimes
|f(x)\oplus b\rangle
\end{eqnarray}
for all $x\in\{0,1\}^n$ and $b\in\{0,1\}$,
where $g$ is a certain function.
Let us define the $(n+1)$-qubit unitary $V$ by
\begin{eqnarray}
V\equiv
(H^{\otimes n}\otimes I)
U^\dagger
(I^{\otimes n}\otimes Z)
U
H^{\otimes n+1}.
\end{eqnarray}
Then,
\begin{eqnarray}
\eta\equiv|\langle0^{n+1}|V|0^{n+1}\rangle|^2
=\frac{gap(f)^2}{2^{2n+1}}.
\end{eqnarray}
From now on the same proof holds as the proof of 
Theorem~\ref{theorem:general:DQC1}
with $\xi=1$.
Therefore, we can construct
the $N$-qubit DQC1 model with $N=n+3$ such that
its acceptance probability
$p_{acc}$ satisfies
$p_{acc}>0$ when $gap(f)\neq0$, and
$p_{acc}=0$ when $gap(f)=0$.
If $p_{acc}$ is classically sampled within a multiplicative
error $\epsilon<1$ in $2^{(1-a)(N-3)}=2^{(1-a)n}$-time,
Conjecture~\ref{conjecture:NC} is violated.

\subsection{Proof of Theorem~\ref{theorem:reversible:DQC1}}
For a given polynomial-size classical reversible circuit 
$C:\{0,1\}^{n+\xi}\to\{0,1\}^{n+\xi}$ that consists of only NOT and TOFFOLI,
its quantum version, $U$, works
as
\begin{eqnarray}
U(|x\rangle\otimes|0^\xi\rangle)=|junk(x)\rangle
\otimes|C_{n+\xi}(x0^\xi)\rangle
\end{eqnarray}
for any $x\in\{0,1\}^n$, where $junk(x)\in\{0,1\}^{n+\xi-1}$
is a bit string (it is actually the first $n+\xi-1$ bits
of $C(x0^\xi)$.)
Therefore, the same proof as that of Theorem~\ref{theorem:general:DQC1}
holds by considering $C_{n+\xi}(x0^\xi)$ as $f(x)$.
Hence we can construct
the $N$-qubit DQC1 model with $N=n+\xi+2$ whose
acceptance probability cannot be classically sampled
within a multiplicative error $\epsilon<1$
in $2^{(1-a)(N-\xi-2)}=2^{(1-a)n}$-time.

\subsection{Proof of Theorem~\ref{theorem:reversible:HC1Q}}
It is the same as the proof of 
Theorem~\ref{theorem:general:HC1Q} by replacing
$f(x)$ with $C_{n+\xi}(x0^\xi)$.

\subsection{Proof of Theorem~\ref{theorem:ETH}}
For a given 3-CNF with $m$ clauses, $f:\{0,1\}^n\to\{0,1\}$, let
us construct the unitary operator $U$ such that
\begin{eqnarray}
U(|x\rangle\otimes|0^\xi\rangle)=|junk(x)\rangle\otimes|f(x)\rangle
\end{eqnarray}
for any $x\in\{0,1\}^n$, where
$junk(x)\in\{0,1\}^{n+\xi-1}$ is a certain bit string,
and $U$ consists of only NOT and TOFFOLI.
Such $U$ is constructed by replacing each AND and OR in $f$ with
TOFFOLI. The 3-CNF $f$ contains $2m$ OR gates and
$m-1$ AND gates.
The replacement of a single AND (or OR) gate with TOFFOLI needs
a single ancilla qubit initialized to 0, and therefore
\begin{eqnarray}
\xi=2m+m-1=3m-1.
\end{eqnarray}
Let us define the circuit $V$ by
\begin{eqnarray}
V\equiv (H^{\otimes n+\xi-1}\otimes X)
U(H^{\otimes n}\otimes I^{\otimes \xi}).
\end{eqnarray}
Then 
\begin{eqnarray}
p_{acc}&\equiv&|\langle 0^{n+\xi}|V|0^{n+\xi}\rangle|^2\nonumber\\
&=&|\langle +^{n+\xi-1}1|\frac{1}{\sqrt{2^n}}\sum_{x\in\{0,1\}^n}
|junk(x)\rangle\otimes|f(x)\rangle|^2\nonumber\\
&=&\frac{(\# f)^2}{2^{2n+\xi-1}}.
\end{eqnarray}
The circuit $V$ has $3m-1$ TOFFOLI gates.
A single TOFFOLI gate can be represented by Clifford gates
and seven $T$ gates~\cite{NC}. Therefore $V$ has $t=21m-7$ $T$ gates,
which means $m=\frac{t+7}{21}$.
Assume that $p_{acc}$ 
is calculated
within an additive error
smaller than $2^{-\frac{3t+14}{7}}$ and
in $2^{o(t)}$ time.
Then, since
\begin{eqnarray}
\frac{1}{2^{\frac{3t+14}{7}}}\le\frac{1}{2}\times\frac{1}{2^{2n+\xi-1}}
\end{eqnarray}
(note that $n\le 3m$),
and   
\begin{eqnarray}
2^{o(t)}=2^{o(\frac{t+7}{21})}=2^{o(m)},
\end{eqnarray}
it means that
$\#f>0$ or $\#f=0$ can be decided in $2^{o(m)}$ time,
which contradicts to Conjecture~\ref{conjecture:ETH}.

\subsection{Proof of Theorem~\ref{theorem:ETHN}}
For a given 3-CNF with $m$ clauses, $f:\{0,1\}^n\to\{0,1\}$, let
us construct the same unitary operator $U$ such that
\begin{eqnarray}
U(|x\rangle\otimes|0^\xi\rangle)=|junk(x)\rangle\otimes|f(x)\rangle
\end{eqnarray}
for any $x\in\{0,1\}^n$, where
$junk(x)\in\{0,1\}^{n+\xi-1}$ is a certain bit string,
$U$ consists of only NOT and TOFFOLI,
and 
$\xi=3m-1$.
Let us define the circuit $V$ by
\begin{eqnarray}
V\equiv H^{\otimes n+\xi}
(I^{\otimes n+\xi-1}\otimes Z)
U(H^{\otimes n}\otimes I^{\otimes \xi}).
\end{eqnarray}
Then 
\begin{eqnarray}
p_{acc}\equiv|\langle 0^{n+\xi}|V|0^{n+\xi}\rangle|^2
=\frac{gap(f)^2}{2^{2n+\xi}}.
\label{citePBC}
\end{eqnarray}
Since $V$ has $t=21m-7$ $T$ gates,
if $p_{acc}$ 
can be classically sampled
within a multiplicative error $\epsilon<1$
in $2^{o(t)}$ time,
it means $gap(f)\neq0$ or $gap(f)=0$ is decided in
non-deterministic $2^{o(m)}$ time, which contradicts to
Conjecture~\ref{conjecture:ETHN}.

\subsection{Proof of Theorem~\ref{theorem:PBC}}
It was shown in Ref.~\cite{BravyiSmithSmolin}
that any quantum circuit over Clifford gates and $t$ $T$ gates
can be simulated by a $t$-qubit PBC.
Therefore,
the acceptance probability $p_{acc}$ of Eq.~(\ref{citePBC})
can be exactly sampled with a $t=21m-7$ qubit PBC.
Due to Conjecture~\ref{conjecture:ETHN}, $p_{acc}$ cannot be
classically sampled within a multiplicative error $\epsilon<1$
and time $2^{o(m)}=2^{o(t)}$.

\subsection{Proof of Theorem~\ref{theorem:H}}
Given a polynomial-size 
Boolean circuit $f:\{0,1\}^n\to\{0,1\}$, construct the unitary
$U$ that consists of only NOT and TOFFOLI such that
\begin{eqnarray}
U(|x\rangle\otimes|0^\xi\rangle)=|junk(x)\rangle\otimes|f(x)\rangle
\end{eqnarray}
for any $x\in\{0,1\}^n$
by replacing each AND and OR of $f$ with TOFFOLI,
where $\xi=poly(n)$ is the number of AND and OR gates
in $f$ and $junk(x)\in\{0,1\}^{n+\xi-1}$
is a certain bit string.
From $U$, we can construct $V$ such that
\begin{eqnarray}
V(|x\rangle\otimes|0^\xi\rangle\otimes|0\rangle)
=|x\rangle\otimes|0^\xi\rangle\otimes|f(x)\rangle
\end{eqnarray}
for any $x\in\{0,1\}^n$.
In fact,
with a CNOT gate, we can copy the value of $f(x)$  
to an additional ancilla qubit as
\begin{eqnarray}
&&|junk(x)\rangle\otimes|f(x)\rangle\otimes|0\rangle\nonumber\\
&\to&
|junk(x)\rangle\otimes|f(x)\rangle\otimes|f(x)\rangle.
\end{eqnarray}
We then apply $U^\dagger$ so that
\begin{eqnarray}
&&U^\dagger(|junk(x)\rangle\otimes|f(x)\rangle)\otimes|f(x)\rangle\nonumber\\
&=&
|x\rangle\otimes|0^\xi\rangle\otimes|f(x)\rangle.
\end{eqnarray}
Then
if we define $W$ by
\begin{eqnarray}
W\equiv
(H^{\otimes n}\otimes I^{\otimes \xi}\otimes H)
V
(H^{\otimes n}\otimes I^{\otimes \xi+1}),
\end{eqnarray}
we obtain
\begin{eqnarray}
|\langle0^{n+\xi}1|
W
|0^{n+\xi+1}\rangle|^2
=\frac{gap(f)^2}{2^{2n+1}},
\end{eqnarray}
which cannot be classically sampled 
within a multiplicative error $\epsilon<1$ in
$2^{(1-a)n}=2^{(1-a)(\frac{h}{2}-\frac{1}{2})}$ time,
where $h\equiv2n+1$ is the number of $H$ in $W$.

\subsection{Proof of Theorem~\ref{theorem:strongH}}
Given a CNF, $f:\{0,1\}^h\to\{0,1\}$, with at most $ch$ clauses,
let us construct the $N\equiv h+\xi+1$ qubit quantum circuit $U$ such that
\begin{eqnarray}
U(|x\rangle\otimes|0^\xi\rangle\otimes|0\rangle)
=|x\rangle\otimes|0^\xi\rangle\otimes
|f(x)\rangle
\end{eqnarray}
for any $x\in\{0,1\}^h$. Such $U$ is constructed
by replacing AND and OR of $f$ with TOFFOLI.
Therefore $U$ consists of $X$, CNOT, and TOFFOLI.
The CNF $f$ contains at most $ch(2h-1)$ OR gates and
$ch-1$ AND gates. 
Therefore, $\xi=ch(2h-1)+ch-1=2ch^2-1$.
If we define the $N$-qubit quantum circuit $V$ by
\begin{eqnarray}
V\equiv U(H^{\otimes h}\otimes I^{\otimes \xi+1}),
\end{eqnarray}
then,
\begin{eqnarray}
\|(I^{\otimes N-1}\otimes|1\rangle\langle1|)V|0^N\rangle\|^2
=\frac{\#f}{2^h}.
\end{eqnarray}
If it is exactly classically calculated in deterministic $2^{(1-a)h}$ time,
we can decide $\#f>0$ or $\#f=0$ in
deterministic $2^{(1-a)h}$ time, which violates SETH.

\subsection{Proof of Theorem~\ref{theorem:tamaki}}
By using a similar construction as that in Ref.~\cite{Dawson},
for any $a,b\in\{0,1\}^N$, $\langle a|U|b\rangle$ can be written as
\begin{eqnarray}
\langle a|U|b\rangle=\frac{1}{\sqrt{2^h}}\sum_{x\in\{0,1\}^h}
(e^{i\frac{\pi}{4}})^{f_1(x)}(-1)^{f_2(x)},
\end{eqnarray}
where $h$ is the number of $H$ gates in $U$,
\begin{eqnarray}
f_1(x)=\sum_{i=1}^h\alpha_ix_i+\beta
\end{eqnarray}
with $\alpha_i,\beta\in\{1,2,3,4,5,6,7\}$, and
\begin{eqnarray}
f_2(x)=\sum_{i=1}^h\sum_{j=1}^h
\gamma_{i,j}x_ix_j+\delta
\end{eqnarray}
with $\gamma_{i,j},\delta\in\{0,1\}$.

For a degree~$2$ polynomial $p \in \mathbb{Z}[x_1,x_2,\ldots,x_h]$ 
and an integer $j \in \{0,1,\ldots,7\}$, 
we define a Boolean function $g_j: \bits{h} \to \bit$ as
\[
g_j(x) =
\begin{cases}
1 & \text{if $p(x) \equiv j \bmod 8$,}\\
0 & \text{otherwise.}
\end{cases}
\]
We have
\begin{eqnarray}
\sum_{x \in \bits{h}} (e^{i\frac{\pi}{4}})^{p(x)}
&=& 
\sum_{x \in \bits{h}} \sum_{j=0}^7 (e^{i\frac{\pi}{4}})^j g_j(x)\\
&=&\sum_{j=0}^7 (e^{i\frac{\pi}{4}})^j  \sum_{x \in \bits{h}} g_j(x).
\end{eqnarray}
We will show that for some $a<1$ and any $j$, 
$\sum_{x \in \bits{h}} g_j(x)$ can be computed in deterministic 
time $2^{ah+o(h)}$.
First we represent $g_j$ as a polynomial in 
$\mathbb{Z}_2[x_1,x_2,\ldots,x_h]$.
\begin{fact}
For each  $j \in \{0,1,\ldots,7\}$, 
there exists a degree~$14$ polynomial $p_j \in \mathbb{Z}[x_1,x_2,\ldots,x_h]$ 
such that $g_j(x) \equiv p_j(x) \bmod 2$ holds for all $x \in \bits{h}$.
\end{fact}

{\it Proof}.
The fact is an immediate consequence of the following.
\begin{lemma}[see e.g. \cite{BeigelT94,ChandraSV84}]
There is a degree $7$ polynomial $q \in \mathbb{Z}[x]$ such that
\[
q(x) =
\begin{cases}
1 & \text{if $x \equiv 0 \bmod 8$,}\\
0 & \text{otherwise,}
\end{cases}
\]
holds for all $x\in \mathbb{Z}$.
\end{lemma}
Setting $p_j(x)=q(p(x)-j)$ completes the proof.\fbox

At this point, computing $\sum_{x \in \bits{h}} g_j(x)$ is reduced to counting the number of roots of the equation $p_j(x) \equiv 1 \bmod 2$.
To do so, we will follow the approach of Lokshtanov 
et al.~\cite{Tamaki}.
We make use of $k$-modulus amplifying polynomials given by the following lemma.
\begin{lemma}[\cite{BeigelT94}]
For all positive integer $k$, there is a degree $2k-1$ polynomial $r_k \in \mathbb{Z}[x]$
such that for all $x \in \mathbb{Z}$, it holds that
\[
\begin{array}{lll}
x \equiv 0 \bmod 2 & \Rightarrow & r_k(x) \equiv 0 \bmod 2^k,\\
x \equiv 1 \bmod 2 & \Rightarrow & r_k(x) \equiv 1 \bmod 2^k.
\end{array}
\]
\end{lemma}

For a positive integer $k$ and an integer $j \in \{0,1,\ldots,7\}$, let
\[
s_{j,k}(y)=\sum_{z \in \bits{k}} r_{k+1}(p_j(y_1,\ldots,y_{h-k},z_1,\ldots,z_k)).
\]
Then we have:
\begin{fact}\label{fact:partial-sum}
For all $y \in \bits{h-k}$,
\[
\sum_{z \in \bits{k}} g_j(y_1,\ldots,y_{h-k},z_1,\ldots,z_k) \equiv
s_{j,k}(y) \bmod 2^{k+1}.
\]
\end{fact}

Since $r_{k+1}(p_j(y_1,\ldots,y_{h-k},z_1,\ldots,z_k))$ is a multi-linear polynomial of degree $14(2k+1)$, 
we can write it down as a sum of terms in time
\[
\binom{h-k}{14(2k+1)} 2^{o(h)}.
\]
for each $z \in \bits{k}$.
Thus, we can write down $s_{j,k}(y)$ as a sum of terms in time
\[
\binom{h-k}{14(2k+1)} 2^{k+o(h)}.
\]

Then, we apply the following lemma.
\begin{lemma}[see e.g. Section 6.2 in \cite{Williams11}]
Given a multi-linear polynomial 
$p \in \mathbb{Z}[x_1,\ldots,x_h]$ represented as a sum of terms,
we can obtain the evaluation of $p$ for all $x\in\{0,1\}^h$
in time $2^{h+o(h)}$.
\end{lemma}
The lemma implies that we can obtain the evaluation of
$s_{j,k}(y)$ for all $y\in\{0,1\}^{h-k}$ in time $2^{h-k+o(h)}$.
By Fact~\ref{fact:partial-sum},  
we also obtain the evaluation of
\begin{eqnarray}
t_j(y)=\sum_{z \in \bits{k}} g_j(y_1,\ldots,y_{h-k},z_1,\ldots,z_k)
\end{eqnarray}
for all $y$.
Finally we obtain the value of $\sum_{x \in \bits{h}} g_j(x)$ 
by calculating $\sum_{y \in \bits{h-k}} t_j(y)$ in time $2^{h-k+o(h)}$.

In the above procedure, required computation time is at most
\[
\binom{h-k}{14(2k+1)} 2^{k+o(h)} + 2^{h-k+o(h)}.
\]
By setting $k=h \times 0.015035\cdots$, both terms are bounded above 
by $2^{0.984965h+o(h)}$.

\subsection{Proof of Lemma~\ref{lemma:noV}}
Assume that a language $L$ is in coNP. Then, if $x\in L$ then
$\#f=0$, and if $x\notin L$ then $\#f>0$, where $f$ is a certain
Boolean circuit.
By the assumption, we can construct the quantum circuit $V$.
Then, if $x\in L$ then $p_{acc}>0$, and
if $x\notin L$ then $p_{acc}=0$.
This means that coNP is in ${\rm NQP}_{\rm DQC1}$,
where ${\rm NQP}_{\rm DQC1}$ is equivalent to NQP except
that the decision quantum circuit is the DQC1 model.
Here NQP is a quantum version of NP, and defined as follows.
\begin{definition}
A language $L$ is in NQP if and only if there exists
a uniformly-generated family $\{V_x\}_x$ of polynomial-size quantum circuits
such that
if $x\in L$ then $p_{acc}>0$ and
if $x\notin L$ then $p_{acc}=0$,
where $p_{acc}$ is the acceptance probability of $V_x$.
\end{definition}
It is known that 
${\rm NQP}={\rm NQP}_{\rm DQC1}$~\cite{Kobayashi,KobayashiICALP},
and therefore we have shown 
${\rm coNP}\subseteq{\rm NQP}={\rm coC}_={\rm P}$.

\subsection{Proof of Theorem~\ref{theorem:stabilizer_rank}}
For a given 3-CNF with $m$ clauses, $f:\{0,1\}^n\to\{0,1\}$,
let us construct the $(n+\xi+1)$-qubit unitary $U$ such that
\begin{eqnarray}
U(|x\rangle\otimes|0^\xi\rangle\otimes|0\rangle)
=|x\rangle\otimes|0^\xi\rangle\otimes|f(x)\rangle
\end{eqnarray}
for any $x\in\{0,1\}^n$, where $\xi=3m-1$.
Such $U$ is constructed as follows.
First, we replace each AND and OR in $f$
with a single TOFFOLI by adding a single ancilla bit initialized to 0. 
Since $f$ contains $2m$ OR gates and $m-1$ AND gates,
the number of TOFFOLI gates (and the number of ancilla bits)
is $\xi=2m+m-1=3m-1$.
Let $U'$ be
thus constructed circuit.
The circuit $U'$ contains NOT gates and $\xi$ TOFFOLI gates.
If we consider it as a quantum unitary circuit,
\begin{eqnarray}
U'(|x\rangle\otimes|0^\xi\rangle)
=|junk(x)\rangle\otimes|f(x)\rangle,
\end{eqnarray}
where $junk(x)\in\{0,1\}^{n+\xi-1}$ is a certain bit string
whose detail is irrelevant here.
Then, we have only to define the $(n+\xi+1)$-qubit unitary $U$
by
\begin{eqnarray}
U\equiv
((U')^{-1}\otimes H)
(I^{\otimes n+\xi-1}\otimes CZ)
(U'\otimes H).
\end{eqnarray}
From the construction, it is clear that
$U$ consists of only $X$, CNOT, and TOFFOLI gates.
Let us define the $(n+\xi+1)$-qubit quantum circuit $V$ by
\begin{eqnarray}
V\equiv 
(H^{\otimes n}\otimes I^{\otimes \xi}\otimes X)
U
(H^{\otimes n}\otimes I^{\otimes \xi+1}).
\end{eqnarray}
Then,
\begin{eqnarray}
\langle0^{n+\xi+1}|V|0^{n+\xi+1}\rangle=
\frac{\#f}{2^n}.
\end{eqnarray}
Let $|\Psi\rangle$ be a resource of
$\xi$ TOFFOLI gates.
Then
there exists an $(n+\xi+1+r)$-qubit Clifford circuit $W$ such that
\begin{eqnarray}
&&\langle0^{n+\xi+1}|V|0^{n+\xi+1}\rangle\nonumber\\
&=&
\eta^{-1}\langle 0^{n+\xi+1+r}|W(|0^{n+\xi+1}\rangle\otimes|\Psi\rangle)
\nonumber\\
&=&\sum_{j=1}^{\chi(|\Psi\rangle)}
\eta^{-1}c_j
\langle 0^{n+\xi+1+r}|W(|0^{n+\xi+1}\rangle\otimes C_j|0^r\rangle),
\label{quantity}
\end{eqnarray}
where
\begin{eqnarray}
\eta\equiv
\Big\|
(I^{\otimes n+\xi+1}\otimes |0^r\rangle\langle0^r|)
|W\Big(|0^{n+\xi+1}\rangle\otimes|\Psi\rangle\Big)\Big\|.
\end{eqnarray}
The quantity of Eq.~(\ref{quantity}) can be
computed in 
$\chi(|\Psi\rangle)\times poly(\xi,n)$
time,
and therefore
\begin{eqnarray}
\chi(|\Psi\rangle)\times poly(\xi,n)\ge 2^{\Omega(m)},
\end{eqnarray}
which means
\begin{eqnarray}
\chi(|\Psi\rangle)\ge
\frac{2^{\Omega(\xi)}}{poly(\xi,n)}
\ge
\frac{2^{\Omega(\xi)}}{poly(\xi)}
=2^{\Omega(\xi)},
\end{eqnarray}
where we have used $n\le 3m=\xi+1$.

\acknowledgements
TM thanks Francois Le Gall, Seiichiro Tani,
and Yuki Takeuchi for discussions.
TM thanks Harumichi Nishimura for discussion and
letting him know Ref.~\cite{Beigel}.
TM thanks Yasuhiro Takahashi for letting him know Ref.~\cite{Cosentino}.
We thank the anonymous referee for valuable comments on our manuscript.
TM is supported by MEXT Q-LEAP, JST PRESTO No.JPMJPR176A,
and the Grant-in-Aid for Young Scientists (B) No.JP17K12637 of JSPS. 
ST is supported by JSPS KAKENHI Grant Numbers 16H02782, 18H04090, 
and 18K11164.


\begin{thebibliography}{99}
\bibitem{noteM}
Note that for the additive error supremacy, all output qubits have to be
measured.
\end{thebibliography}
\end{document}